\documentclass{elsart}%
\usepackage{amsmath}
\usepackage{graphicx}%
\usepackage{amsfonts}%
\usepackage{amssymb}
\begin{document}
\begin{frontmatter}
\title{Finite-Bandwidth Calculations for \\
Charge Carrier Mobility in Organic Crystals}
\author{V.M. Kenkre} \address{Department of Physics and
Astronomy, Center for Advanced Studies,  University of New Mexico,
Albuquerque, NM 87131, USA}

\begin{abstract}
Finite-bandwidth effects on the temperature dependence of the mobility
of injected carriers in pure organic crystals are explored for a 
simplifed case of impurity scattering. Temperature-dependent bandwidth effects
are discussed briefly through a simplified combination of band and polaronic
concepts.
\end{abstract}
\end{frontmatter}  

\section{Introduction}

In studying charge carrier (electron or hole) transport in organic crystals,
one avoids the use of standard band theoretic approaches \cite{ziman} since
the carrier bandwidth is not an overwhelmingly large energy in organic, in
contrast to inorganic, materials \cite{pope}. Mean free paths calculated on
the basis of bare band-theoretic descriptions have been explicitly found to
be smaller than a lattice constant \cite{adk}. Such findings have
necessitated polaronic or hopping theories of transport in organic materials 
\cite{silbeymunn,kadd}. The question arises whether this trend \cite
{schein,kd} away from bare-band theory, which has continued in the
literature on organic materials in recent times \cite{pkd}, needs to be
revised or at least modified in light of recent observations \cite{batlogg}
and calculations \cite{Bredas} on pentacene. The large low-temperature
mobility in these materials (orders of magnitude larger than values reported
earlier \cite{ScheinKarlDuke} at other temperatures and in other organics),
sharp power-law dependence of the mobility on temperature in a wide range 
\cite{batlogg,karl}, and nonlinear saturation phenomena \cite{kp} which have
been interpreted \cite{batlogg} in terms of large bandwidths, all could
point to the need for bare-band theory to be considered seriously in organic
materials. On the other hand, the bandwidths are believed \cite
{batlogg,Bredas} not to be of the order of several eV's as in inorganic
materials but smaller. One concludes, therefore, that required is a
description valid for \emph{intermediate} bandwidths ($B$). To begin the
construction of such a description is the main purpose of the present Note.

Temperature-dependent (Huang-Rhys) $B$'s arise in polaronic transport \cite
{silbeymunn,kadd,holstein}. In the past, effective bare $B$'s, which undergo
reduction by polaronic effects in the presence of strong interactions with
vibrations, were themselves believed (or shown) to be rather small. With
naphthalene as an example, $B\sim $ $10$ eV, the polaronically reduced $B$
is about a tenth of this value in light of the magnitude of the coupling
constants and is thus well surpassed by the thermal energy $k_{B}T$ . The
latter exceeds $2.5$ meV corresponding to $30$ K, the lowest temperature in
the naphthalene observations \cite{ScheinKarlDuke}. Therefore, although past
explanations of mobility observations have certainly made use of the
temperature dependence of polaronic bandwidths \cite{silbeymunn,kadd} to
address decreasing mobility with increasing temperature \cite{kadd},
polaronic calculations have served as the investigational procedure. Band
considerations have either not been used at all \cite{kadd} or assigned a
less important role \cite{silbeymunn}. Estimates in pentacene raise the
possibility that $k_{B}T$ is smaller than $B$ for a part of the
observations. To show the results of a simple combination capable of
blending finite-band theory and polaronic concepts, is a secondary purpose
of the present Note.

\section{Constant Scattering Rates and Truncated Parabolic Bands}

Band-theoretic descriptions in inorganic materials effectively take the
bandwidth $B$ to be infinite. The well-known textbook formula for the
mobility
\begin{equation}
\mu =\left( \frac{q}{k_{B}T}\right) \frac{\int_{0}^{\infty }d\varepsilon
\;v^{2}\left( \varepsilon \right) \tau \left( \varepsilon \right) \rho
\left( \varepsilon \right) e^{-\varepsilon /k_{B}T}}{\int_{0}^{\infty %
}d\varepsilon \;\rho \left( \varepsilon \right) e^{-\varepsilon /k_{B}T}}
\label{mu}
\end{equation}
where $k_{B}$ is the Boltzmann constant, and $q$, $\varepsilon$, $T$, $v$,
 $\rho$, $\tau$ are the carrier charge, energy, temperature, velocity,
density of states, and relaxation time, respectively, has $\infty $ as the
upper limit of the energy integrations. We will explore, instead, the
consequences of replacing it by the finite value of $B$, suitably modifying
the density of states. We consider a one-band model, which means merely that
the band gap is taken to be large enough to make interband transitions
unimportant, and  assume a simple  $\rho \left( \varepsilon \right) $ having
the free-electronic form within the band and to vanish outside: 
\begin{equation}
\rho \left( \varepsilon \right) =N\left( \frac{3\sqrt{\varepsilon }}{2B\sqrt{%
B}}\right) \left[ \theta \left( \varepsilon \right) -\theta \left(
\varepsilon -B\right) \right] 
\end{equation}
where $\theta $ is the Heaviside step function and $N$ is the number of
sites. It is easy to show that the carrier velocity $v\left( \varepsilon
\right) $ is 
\begin{equation}
v=\frac{1}{\hbar }\frac{d\varepsilon }{dk}=C\sqrt{B\varepsilon }
\end{equation}
The constant $C$ equals $\left( 2/\sqrt{3}\right) \left( a/\hbar \right)
\left( 3\pi ^{2}\right) ^{-1/3}$ if the system is isotropic in three
dimensions, $a$ being the lattice constant, given that the charge carriers
are Fermionic ($2$ spin values per orbital state). The thermal distribution
in (\ref{mu}) is classical for the standard reasons that the carriers being
injected are so small in number, e.g., in time-of-flight observations, that
the Fermi distribution is well approximated by a Maxwell distribution.
Although space-charge limited conditions of measurement are common now for
some experimental setups such as in field-effect-transistor geometry, we
will assume here that this assumption is valid. If the need arises, it can
always be relaxed in subsequent calculations.

Our present purpose being to get at the simplest consequences of finite $B$
as economically as possible, for most of the calculations here, we will
first take the relaxation time $\tau $ to have no energy dependence. Such a
case appears in an early analysis of Erginsoy \cite{erg}. Whether the entire
set of assumptions behind Erginsoy's analysis applies to organic crystals is
of no importance in the present context. Our interest lies only in
exploring, in the first instance, the consequences of a constant $\tau $ in
a finite band system. The finite-band generalization of (\ref{mu}) is then
\begin{equation}
\mu ^{\prime }=\frac{\mu }{\hbar qC^{2}}=\left( \frac{B}{S}\right) \frac{%
\int_{0}^{B/k_{B}T}dx\;x\sqrt{x}e^{-x}}{\int_{0}^{B/k_{B}T}dx\;\sqrt{x}e^{-x}%
}=\left( \frac{B}{S}\right) \left( \frac{B}{k_{B}T}\right) \frac{%
\int_{0}^{1}dx\;x\sqrt{x}e^{-xB/k_{B}T}}{\int_{0}^{1}dx\;\sqrt{x}%
e^{-xB/k_{B}T}}.  \label{twomu}
\end{equation}
Displayed in (\ref{twomu}), is the dimensionless ratio $\mu ^{\prime }$ of
the mobility to $\hbar qC^{2}=\left( 4a^{2}q/3\hbar \right) \left( 3\pi %
^{2}\right) ^{-2/3}$ , and $S$ is the `scattering energy' $\hbar /\tau $.
The dimensionless mobility $\mu ^{\prime }$ has been expressed in terms of
the dimensionless ratios of the bandwidth to the scattering energy and to
the thermal energy: $B/S$ and $B/k_{B}T$. While both expressions in (\ref
{twomu}) are exact, they can be used respectively in opposite limits most
conveniently:
\begin{eqnarray}
\mu ^{\prime }&=&\left( 3/2\right) \left( B/S\right) \mbox{ for }%
B/k_{B}T\rightarrow \infty \nonumber \\
\mu ^{\prime }&=&\left(%
3/5\right) \left( B/S\right) \left( B/k_{B}T\right) \mbox{ for }%
B/k_{B}T\rightarrow 0.  \label{bs}
\end{eqnarray}
This $B$ to $B^{2}$ transition as one goes from large bands to small bands
arises from the occupation of states in the band. The mobility, which
involves the thermal average of the square of the velocity, is proportional
to $Bk_{B}T$, the product of $B$ and the thermal energy in the large
bandwidth case, but to $B^{2}$ in the small bandwidth case, because the
temperature is so large with respect to $B/k_{B}$ that it makes no
contribution: all states in the band are occupied equally. The transition
takes on different character if the relaxation time $\tau $ is
energy-dependent, the limiting exponents of $B$ being different from $1$ and 
$2$ respectively.

The two different ways of scaling the energy apparent in (\ref{twomu}) are
particular cases $Q=k_{B}T$ and $Q=B$ of scaling with an arbitrary quantity $%
Q$ having the dimensions of energy: 
\begin{equation}
\mu ^{\prime }=\left( \frac{B}{S}\right) \left( \frac{Q}{k_{B}T}\right) 
\frac{\int_{0}^{B/Q}dx\;x\sqrt{x}e^{-xQ/k_{B}T}}{\int_{0}^{B/Q}dx\;\sqrt{x}%
e^{-xQ/k_{B}T}}  \label{Qeqn}
\end{equation}
This general expression can be used in additional ways such as by taking $%
Q=S:$%
\begin{equation}
\mu ^{\prime }=\left( \frac{B}{k_{B}T}\right) \frac{\int_{0}^{B/S}dx\;x\sqrt{%
x}e^{-xS/k_{B}T}}{\int_{0}^{B/S}dx\;\sqrt{x}e^{-xS/k_{B}T}}.
\end{equation}
In that case, extreme limits of $B/S$ can be studied---if it is very large,
which represents the large mean free path case, $\mu ^{\prime }$ is
proportional to a ratio of Laplace transforms of powers, with $S/k_{B}T$ as
the Laplace variable.

The three representations of the mobility given above are useful in
understanding limiting behavior of carrier transport in extremes of the
three respective ratios of $B,$  $k_{B}T$, and  $S$. Generally, one can
identify the integrals in (\ref{twomu}) with incomplete gamma functions $%
\gamma $ defined via $\gamma \left( b,x\right) =\int_{0}^{x}e^{-t}t^{b-1}dt,$
and write, exactly, i.e., for arbitrary relative values of $B,$ $k_{B}T,$
and $S$, 
\begin{equation}
\mu ^{\prime }=\left( \frac{B}{S}\right) \frac{\gamma \left(
5/2,B/k_{B}T\right) }{\gamma \left( 3/2,B/k_{B}T\right) }  \label{gammamu}
\end{equation}
which can easily be shown to lead to the various respective limits, or
rewritten in other ways such as in terms only of error functions and
exponentials through the use of the chain condition $\gamma \left(
b+1,x\right) =b\gamma \left( b,x\right) -x^{b}e^{-x}$ and the relation $%
\gamma \left( 1/2,x\right) =\sqrt{\pi }{\,\mbox{erf}}\left( \sqrt{x}\right).$

\section{Power Laws}

The sharpness of the power laws in the temperature dependence of the
observed $\mu $ in organic crystals, whether in the recent experiments on
pentacene \cite{batlogg} or in experiments on other crystals such as
naphthalene reported a couple of decades back \cite{karl}, is impressive.
Although Giuggioli et al. \cite{lucaetc} have shown recently that visually
acceptable fits to the data can be produced by a band theory addressing the
partial range of pentacene data from 20-400 K by combining acoustic and
phonon scattering, it is interesting to ask if a true power law form can be
obtained as an \emph{analytic limit} from the expressions. The following
simplified approach produces such as a limit. Assume that the mechanism of
scattering is such that the relaxation time $\tau $ depends both on the
carrier energy $\varepsilon $ and the temperature $T$ as  $\tau =\tau
_{0}\left( \varepsilon /\varepsilon _{0}\right) ^{-p}\left( T/T_{0}\right)
^{-r}$ where $\varepsilon _{0}$, $\tau _{0}$, and $T_{0}$ are a
characteristic energy, time, and temperature respectively. Such expressions
arise naturally in many contexts. Thus, $p=1/2$ in many semiconductors \cite
{Conwellbook}, and $r=1$ for acoustic phonon scattering at high
temperatures. Equation (\ref{mu}) then yields 
\begin{equation}
\mu ^{\prime }=\left( \frac{B\tau _{0}}{\hbar }\right) \left( \frac{%
\varepsilon _{0}}{k_{B}T_{0}}\right) ^{p}\frac{\gamma \left(
5/2-p,B/k_{B}T\right) }{\gamma \left( 3/2,B/k_{B}T\right) }\left( \frac{T}{%
T_{0}}\right) ^{-\left( p+r\right) }  \label{powermu}
\end{equation}
Whether or not this appears as a power form depends on the particular $T$
dependence of the ratio of the $\gamma $-functions. For $p=0$, one recovers (%
\ref{gammamu}) except for the explicit $T^{-r}$ factor. For $p=1$, the $%
\gamma $-functions cancel and the mobility is seen to display a clear
temperature law $T^{-\left( 1+r\right) }.$ In such a case the value $r=1.7\ $%
could conceivably correspond to the observed $T^{-2.7}$ dependence in
pentacene \cite{batlogg}. It is easy to see that, even if the value of $p$
is not $1$, a sharp power law may indeed be the direct apparent consequence
of the assumed $\tau$ in an appropriate (although limited) range of temperature,
provided the ratio of the $\gamma $-functions in (\ref{powermu}) is largely $%
T$-independent in that range.

\section{Huang-Rhys Bandwidths and Polaronic Expressions}

If polaronic effects are present in carrier transport, it could be argued 
\cite{batlogg} that some of the physics might be captured by invoking the
well-known Huang-Rhys dependence 
\begin{equation}
B=\widetilde{B}=B_{0}e^{-G^{2}\coth \left( \hbar \Omega /2k_{B}T\right) }
\label{hr}
\end{equation}
where $B_{0}$ is the bare bandwidth, and its exponential reduction occurs as
a result of strong interaction via the coupling constant $G$ with vibrations
of frequency $\Omega $ . For instance, the substitution of (\ref{hr}) in the
simple constant-$\tau $ version (\ref{gammamu}) would yield the normalized
mobility
\begin{equation}
\frac{\mu \left( T\right) }{\mu (0)}=\frac{e^{-G^{2}\left[ \coth \left( 
\frac{\hbar \Omega }{2k_{B}T}\right) -1\right] }}{3/2}\left[ \frac{\gamma
\left( \frac{5}{2},\frac{\widetilde{B}}{k_{B}T}\right) }{\gamma \left( \frac{%
3}{2},\frac{\widetilde{B}}{k_{B}T}\right) }\right] .  \label{adhoc}
\end{equation}
Equation (\ref{adhoc}) displays the qualitative trend of the low temperature
data, but not of the high temperature data, in the acenes. Furthermore,
there is always a large enough temperature at which the Huang-Rhys factors
reduce the polaron bandwidth enough to make the mean free path smaller than
a lattice constant. This forces the band calculation to lose its accuracy
and, indeed, its applicability. Crystal momentum becomes an inadequate
quantum number in such a case, and a hopping description becomes necessary.
In particular, a bridging of the present finite-band calculations and of
polaronic theories developed earlier \cite{silbeymunn,kadd} is required. How
would one carry out such a combination? This important and difficult
question, along with application to pentacene observations, is being
addressed in our ongoing work. Here we sketch a simplified answer to the
question.

Polaronic mobility theory for narrow bands has been successfully applied 
\cite{kadd} more than a decade ago to naphthalene observations \cite
{ScheinKarlDuke} on the basis of an expression, which, in one of its most
simplified forms, may be written as 
\begin{equation}
\mu =\frac{cq}{k_{B}T}\left( \frac{\widetilde{B}a}{\hbar }\right) ^{2}\left( 
\frac{1}{\alpha }\right) I_{0}\left( \frac{2G^{2}}{\sinh \left( \frac{\hbar 
\Omega }{2k_{B}T}\right) }\right)   \label{polaron}
\end{equation}
where $c$ is a numerical constant and $I_{0}$ is a modified Bessel function.
The scattering rate $\alpha $ is essentially the constant $1/\tau $
appearing in band expressions shown earlier in this Note. The expression
comes about from the time integration of the velocity autocorrelation
function$.$ The velocity autocorrelation function is a product of a polaron
part which arises from the interaction of the charge carrier with vibrations
and a part $e^{-\alpha t}$ which arises from interaction with static defects 
\cite{vmk}. Equation (\ref{polaron}) is appropriate to a narrow band. An
extension of the narrow-band expression to an intermediate-band situation
may be made by using the calculation given earlier in this Note to obtain
the static part $e^{-\alpha t}$ . It leads to the replacement of $\alpha $
in (\ref{polaron}) by an effective $\alpha $ which is dependent on $T$ and $B
$ in addition to being proportional to the constant $1/\tau $. The result is
(with $c^{\prime }$ a numerical constant) 
\begin{equation}
\mu =c^{\prime }q\tau \left( \frac{a}{\hbar }\right) ^{2}\widetilde{B}\left[ 
\frac{\gamma \left( \frac{5}{2},\frac{\widetilde{B}}{k_{B}T}\right) }{\gamma
\left( \frac{3}{2},\frac{\widetilde{B}}{k_{B}T}\right) }\right] I_{0}\left( 
\frac{2G^{2}}{\sinh \left( \frac{\hbar \Omega }{2k_{B}T}\right) }\right) 
\label{grand}
\end{equation}
The general result (\ref{grand}) combines the polaronic and band-theoretic
character in a simple way. It reduces to the polaronic form (\ref{polaron})
for large $T$, i.e., for $k_{B}T>>\widetilde{B}$ where $\widetilde{B}$ is
the reduced bandwidth given by (\ref{hr}). In this limit, the reduced band
is fully occupied and the ratio of the two $\gamma $-functions simplifies to 
$\left( 3/5\right) (\widetilde{B}/k_{B}T).$ In the opposite limit of small
temperatures, $k_{B}T<<\widetilde{B}$, the occupation of the band is
controlled by the temperature, the zero-$T$ limit of the ratio of the $%
\gamma $-functions is simply $3/2,$ and (\ref{grand}) simplifies to the band
result. Indeed, if also $k_{B}T<<$ $\hbar \Omega /2$, the argument of the $I$%
-Bessel function is negligible and the Bessel function is simply $1.$ The
mobility then reduces to (\ref{bs}) with the zero-$T$ reduced bandwidth $%
B_{0}e^{-G^{2}}$ in place of $B$ in (\ref{bs}). Generally, the mobility,
normalized to its $T=0$ value, is 
\begin{equation}
\frac{\mu \left( T\right) }{\mu (0)}=\mu _{ah}\left( T\right) I_{0}\left( 
\frac{2G^{2}}{\sinh \left( \frac{\hbar \Omega }{2k_{B}T}\right) }\right) 
\label{last}
\end{equation}
where $\mu _{ah}\left( T\right) $ is the right hand side of (\ref{adhoc}).

The \emph{ad hoc} expression (\ref{adhoc}) obtained by straightforward
substitution of the bare bandwidth $B$ by the polaronic reduced bandwidth $%
\widetilde{B}$ in (\ref{gammamu}) differs precisely by the introduction in (%
\ref{last}) of the Bessel function factor. This agrees with the well known
fact \cite{silbeymunn,kadd} that the excess of that factor over $1$
represents the hopping contribution of the mobility. We plot (\ref{last})
in Fig. 1. Parameters chosen are arbitrary but illustrate the respective
behaviors reported in the literature for naphthalene where a flat
temperature dependence is observed \cite{ScheinKarlDuke}, and for pentacene
where a turn-over with increasing mobility in the higher temperature region
dependence has been reported \cite{batlogg}. 
\begin{figure}
\centering
\resizebox{\columnwidth}{!}{\includegraphics{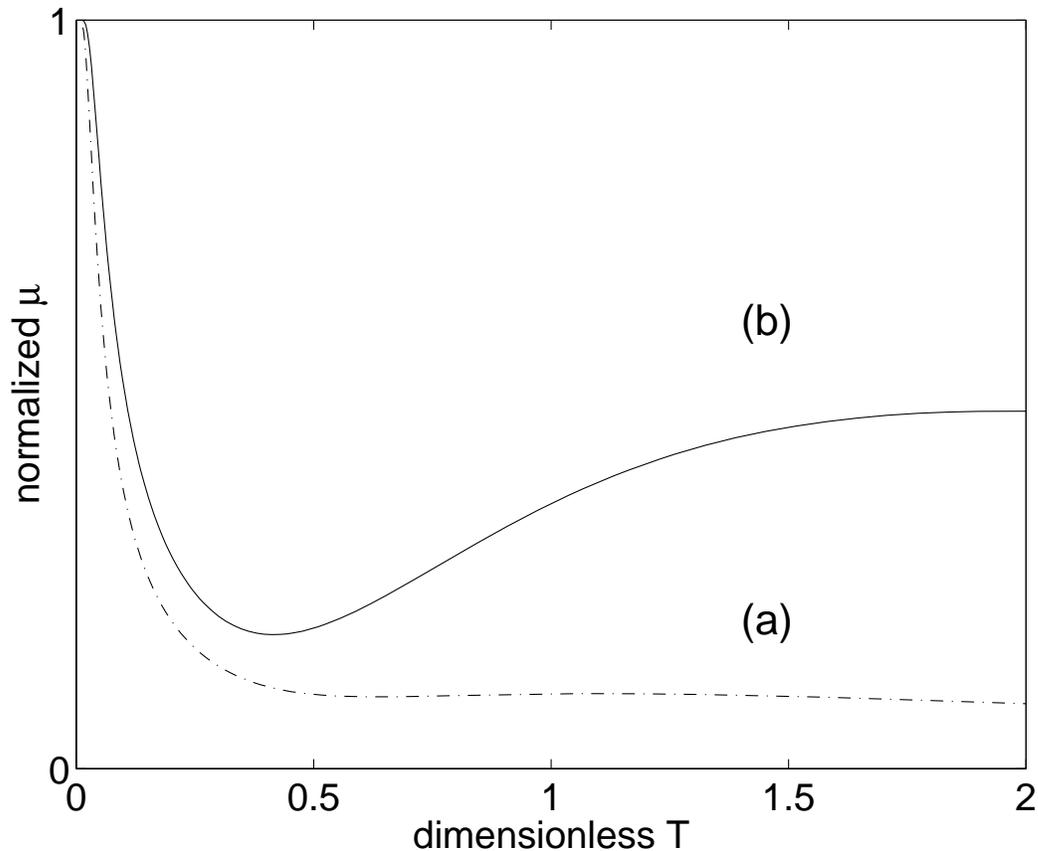}}
\caption{Dependence of the mobility $\mu $ on the temperature $T$ plotted
for two (arbitrary) parameter combinations. In units of $\hbar \Omega $, the
bare bandwidth $B_{0}$ is 0.5 and the coupling constant $g$ is 1.5 in (a)
while $B_{0}$ is 2 and $g$ is 1.8 in (b). The temperature $T$ is plotted in
units of $\hbar \Omega /2k_{B}$ and $\mu $ is normalized to its value at $%
T=0.$ Qualitatively only, (a) and (b) have resemblance to naphthalene and
pentacene reported data respectively.}
\end{figure}

\section{Remarks}

Through simplified calculations, we have presented here arguments towards an
intermediate finite-bandwidth theory which extends previous narrow-band
treatments. The new results obtained  are: finite-band effects stemming from
full occupation of the band at temperatures large enough so that the thermal
energy exceeds the bandwidth, $B$ to $B^{2}$ transition of the mobility
dependence on the bandwidth with variation of temperature, an indication of
how power laws might arise as analytical limits of appropriate scattering
mechanisms, the specific form of the mobility under the \emph{ad hoc}
Huang-Rhys temperature dependence of the bandwidths and the relation of such
an expression to polaronic expressions, and a usable interpolation formula
which combines polaronic and bare-band characters.

These calculations should by no means be treated as a complete theory of
polaronic and band effects. More realistic analyses of the band transport
are being carried out by Giuggioli et al. \cite{lucaetc}, taking into
account acoustic and optical phonon interactions which lead to striking new
finite-band effects. While not totally accurate, the simple truncated
free-carrier density of states used here has the correct broad (bare) band
limit in contrast to Gaussian forms \cite{silbeymunn} and should, therefore,
be able to capture the essential physics at that end. It is hoped that the
present simple calculations will contribute towards the construction of a
general picture of quasiparticle transport in pure organic crystals which
treats both the temperature dependence of the mobility, and high field
effects such as velocity saturation that have been treated recently \cite{kp}. 

\ack
It is a pleasure to acknowledge conversations with John Andersen, Luca
Giuggioli and Paul Parris. This research was supported in part by the NSF
under grant DMR-0097204 and by the Los Alamos National Laboratory under a
contract to the Consortium of the Americas for Interdisciplinary Science of
the University of New Mexico.

\end{document}